\documentstyle[psfig,twocolumn,prl,aps,amssymb]{revtex}
\begin{document}
\wideabs{
\draft

\title{Partially spin polarized quantum Hall effect in the
filling factor range $1/3<\nu<2/5$}
\author{Chia-Chen Chang, Sudhansu S. Mandal, and Jainendra K. Jain}
\address{Department of Physics, 104 Davey Laboratory,
The Pennsylvania State University,
Pennsylvania 16802}
\date{\today}

\maketitle

\begin{abstract}
The residual interaction between composite fermions (CFs) can express itself  
through higher order fractional Hall effect.  With the help of diagonalization 
in a truncated composite fermion basis of low-energy many-body states,
we predict that quantum Hall effect with partial spin polarization is  
possible at several fractions between $\nu=1/3$ and $\nu=2/5$.
The estimated excitation gaps are approximately two orders of magnitude smaller
than the gap at $\nu=1/3$, confirming that the inter-CF
interaction is extremely weak in higher CF levels.
\end{abstract}

}


In recent years, the physics arising from the interactions between composite
fermions has come into focus \cite{Pan,Pinczuk,Mandal,Park2,Quinn,Lee}.  
The model of {\em non}interacting
composite fermions explains the fractional quantization of Hall resistance 
\cite{Tsui} at $R_H = h/ f e^2$
with
\begin{equation}
f = \frac{n}{2pn \pm 1}
\end{equation}
as the integral quantum Hall effect \cite{Klitzing} (IQHE) of composite fermions \cite{Jain}.
(Particle hole symmetry in the lowest Landau level implies fractional Hall 
effect also at $1-f$ or $2-f$, for fully or partially spin polarized systems, respectively.) 
The weak residual interaction between composite fermions is often
masked by disorder or temperature, much as the fractional quantum Hall effect (FQHE), a
manifestation of inter-electron interactions, is absent in low mobility samples
or at high temperatures.  However, with improvements in experimental conditions,
the physics originating from the
interaction between composite fermions is beginning 
to emerge \cite{Pan,Pinczuk}.

A possible manifestation of the interaction between composite fermions
will be the appearance of higher order FQHE
states in between the above fractions.
The possible new fractions can be derived straightforwardly \cite{Mandal,Park2,Jain}.
At the {\em non-integral} values of the CF filling factor given by  
\begin{equation}
\nu^*=n + \frac{m}{2{p'}m\pm 1}
\end{equation}
the composite fermions in the topmost partially filled level may capture, as a result of
the interaction between them, $2p'$ additional vortices to transform
into higher order composite fermions and fill $m$ CF levels to 
exhibit new incompressible states, which will produce FQHE 
at $\nu=\frac{\nu^*}{2p\nu^*\pm 1}$
between the fractions $\nu=\frac{n}{2pn\pm 1}$ and
$\nu=\frac{n+1}{2p(n+1)\pm 1}$.
The situation is analogous to the appearance of FQHE of electrons
in partially filled {\em electronic} Landau levels.

Pan {\em et al.} \cite{Pan} have reported FQHE at $\nu=4/11$,  $\nu=5/13$ and $\nu=6/17$ in 
the filling factor range $2/5 > \nu > 1/3$.  The $\nu=4/11$ minimum is seen in 
magnetic fields as high as $B=33$T, pointing to a fully polarized QHE state.
The simplest fractions in the above scenario are $\nu^*=1+1/3$ and $\nu^*=1+2/3$,
which produce higher order QHE at $\nu=4/11$ and
$\nu=5/13$, and $\nu=6/17$ originates from $\nu^*=1+1/5$.  While it is encouraging 
that the desired fractions are obtained, more detailed theoretical investigations 
do not find new FQHE states for these fractions for {\em fully} spin-polarized 
composite fermions for an idealized model
neglecting disorder, transverse thickness, and Landau level mixing\cite{Mandal,Quinn,Lee};
the residual interaction between composite
fermions in {\em higher} CF levels does not appear to be sufficiently strongly 
repulsive at short distances to cause additional vortices to bind to composite fermions.
It is not known at the present which of the neglected effects is responsible for 
the discrepancy.

While only fully spin polarized states are possible at sufficiently high magnetic
fields, where a spin in the wrong direction costs a prohibitively high energy,
FQHE states at the fractions of Eq.~(1) with partial polarizations have been observed
experimentally,  and transitions between
differently polarized states have been studied as a function of the
Zeeman energy\cite{Spin_ex}.
These studies are satisfactorily described by the composite fermion theory
including spin\cite{Wu,Park}.
Composite fermions with spin are analogous to electrons with spin at $\nu^*$, 
with the same Zeeman energy but with an effective cyclotron energy.  
While the Zeeman energy is very small compared to the cyclotron energy for 
electrons (in GaAs), the two are comparable for composite fermions, thus producing a 
richer variety of states\cite{Spin_ex}.
This raises the question of whether FQHE states with partial spin polarization
are possible at fractions like $\nu=4/11$, which is the subject of this paper.
It is stressed that the present work does not purport to be an explanation for the 
observations in Ref.~\onlinecite{Pan}, but a prediction for 
sufficiently low magnetic fields.

Following the standard approach, we will neglect in this study 
the effects of finite thickness and Landau level mixing.
The most reliable method is exact numerical diagonalization, which
is not an option for this problem because of the rather large Hilbert
space.  In this article, we carry out diagonalization in
a truncated low-energy CF basis of many body states \cite{Mandal}.  We will  
concentrate here on the filling factor range $2/5 > \nu > 1/3$,  that is, 
$2>\nu^*>1$.  New FQHE is most likely at $\nu^*=1 + \frac{m}{2m+1}$, which  
correspond to electron filling factors 
$\nu=\frac{3m+1}{8m+3}$.  The positive and negative integral values for 
$m$ produce $\nu= 4/11,\; 7/19, ... $ and $\nu=5/13,\; 8/21, ...$
It will be assumed that
the up-spin composite fermions fill one level
completely, and the down-spin composite fermions have filling
factor $\nu^*_{\downarrow}=m/(2m+1)$ in the lowest spin reversed band,
giving the total CF filling $\nu^*=1+\nu^*_{\downarrow}$.
For our truncated basis, we consider wave functions of the form:
\begin{equation}
\Psi^{\alpha}_{\frac{3m+1}{8m+3}} =
\Phi_1^2 [\Phi_{1,\uparrow} \Phi_{\frac{m}{2m+1},\downarrow}^{\alpha}]
\label{basis}
\end{equation}
Here, $\Phi_{1,\uparrow}$ is the fully occupied up-spin lowest Landau
level band.  $\Phi_{\frac{m}{2m+1},\downarrow}^{\alpha}$ are various orthogonal 
wave functions (labeled by $\alpha$) at filling $\nu^*_\downarrow=\frac{m}{2m+1}$
in the down-spin band, obtained by exact diagonalization at $\nu^*_\downarrow$.  
(Coupling to higher Landau levels is neglected.)
The fully antisymmetric function $\Phi_1$ is one filled 
Landau level of ``spinless" electrons; the Jastrow factor $\Phi_1^2$, 
as always, converts, through attachment of two vortices to each
electron, the $\nu^*=1+\frac{m}{2m+1}$ wave function of electrons
in square brackets to the $\nu^*=1+\frac{m}{2m+1}$
wave function of composite fermions, which is identified with a 
basis function for interacting electrons at $\nu=\frac{3m+1}{8m+3}$.
(The spin part of the wave function is not explicitly shown.  The full
wave function is obtained by multiplying by the spin part
$u_{1}...u_{N_{\uparrow}}d_{N_{\uparrow}+1}...d_N$, where $N_{\uparrow}=
N-N_{\downarrow}$ is the number of up-spin electrons,
followed by antisymmetrization.)
We will study below $m=1$ and $m=2$ ($\nu=4/11$ and $\nu=7/19$).
The states of the form given in Eq.~(\ref{basis}) 
obviously do not exhaust the entire Hilbert
space at $\nu$, as they neglect the mixing between the Landau levels of composite 
fermions, but we believe that they span the low-energy Hilbert space.
If the system is incompressible, the ground state at $\nu$ is likely to be well
described by
\begin{equation}
\Psi^{gr}_{\nu} =
\Phi_1^2 [\Phi_{1,\uparrow} \Phi^{gr}_{\nu^*_{\downarrow},\downarrow}]
\label{incom}
\end{equation}
where $\Phi^{gr}_{\nu^*_{\downarrow},\downarrow}$ is the Coulomb ground 
state at $\nu^*_{\downarrow}$.  For $\nu=\frac{3m+1}{8m+3}$, the 
state at $\nu^*_{\downarrow}=\frac{m}{2m+1}$ is accurately given by  
the standard wave function ${\cal P} \Phi_{1,\downarrow}^2\Phi_{m,\downarrow}$,
where ${\cal P}$ denotes projection into the lowest Landau level (LLL).

It is noteworthy that no assumption
is made regarding the nature of the state in the reversed-spin
sector, and the calculation can in principle give either a
compressible or an incompressible ground state.
Indeed, a similar study for {\em fully} spin polarized
systems at many fractions like $\nu=4/11$ failed to yield an incompressible
ground state \cite{Mandal}, contrary to what one might have naively expected.

An earlier study\cite{Park2} began with the {\em assumption} of
the partially spin polarized ground state
\begin{equation}
\Psi^{gr}_{\nu=4/11} = \Phi_1^2 \Phi_{1,\uparrow} [\Phi_{1,\downarrow}]^3
\end{equation}
which is derived from Laughlin's wave function\cite{Laughlin} 
$[\Phi_{1,\downarrow}]^3$ in the spin reversed sector \cite{Halperin},  
and considered a trial wave function for its neutral
excitation containing a pair of CF particle hole pair in the
reversed spin sector.   It was found that the energy of the excitation
remains positive for all wave vectors, indicating that the assumed
ground state wave function is stable against excitations.
While this study did not eliminate FQHE at $\nu=4/11$, it did not
test whether the ground state is necessarily incompressible, and if so,
whether it is well described by the trial wave function in Eq.~(\ref{incom}).
The present study provides a more rigorous (though still not conclusive)
test for partially polarized QHE at $\nu=4/11$.

In the following discussion we will employ the spherical geometry \cite{Monopole,JK}, 
where we consider $N$ electrons moving on the surface of a sphere at the presence
of a magnetic monopole with strength $Q$ at the center.
The magnitude of the radial magnetic field $B$ is given by
$2Q\phi_0/4\pi R^2$, where
$\phi_0 = he/c$ is the flux quantum, $R$ is the radius of the sphere, and
$Q$ is either an integer or a half-integer due to the Dirac quantization condition.
The composite fermion theory maps the system of interacting electrons
at $Q$ to the CF system at $q^*=Q-p(N-1)$. It is convenient to label the wave function
by the monopole strength; for example,
the wave function at $Q$ is obtained from the
electron wave function $\Phi_{q^*}$ by
$\Psi_{Q} = {\cal P}\,[{\Phi_1}^{2p}\,\Phi_{q^*}]$.

The CF theory fixes the relation between $N$ and $Q$ as follows.
The effective $q^*$ is determined by requiring that $N_{\downarrow}$
electrons have filling $\nu^*_{\downarrow}=\frac{m}{2m+1}$:
$q^*=N_{\downarrow}(2m+1)/2m-(m+2)/2$.
With $N_{\uparrow}=2q^*+1$ and $Q=q^*-(N-1)$, we get
\begin{equation}
 Q=\frac{8m+3}{6m+2}N-\frac{m^2+10 m+3}{6m+2}.
\end{equation}
Therefore, at $\nu=4/11$, where the effective filling $\nu^*=1+1/3$ with $m=1$, the
relation is
\begin{equation}
Q_{4/11}=(11 N-14)/8.
\end{equation}
Similarly, for $\nu=7/19$, where $\nu^*=1+2/5$ and $m=2$,
\begin{equation}
Q_{7/19}=(19 N-27)/14.
\end{equation}
Note that both relations give the desired filling factors in the thermodynamic limits:
$\nu=\lim_{N\rightarrow\infty}\frac{N}{2Q}$.
For a given particle number $N$, the pair $(N,Q)$ is the only input in the numerical
calculation. Table~I gives the systems we have studied below.

The energy spectrum is calculated numerically by the Monte Carlo (MC) method, 
following Ref.~\onlinecite{Mandal}.
Because the low energy basis states from exact diagonalization
are not necessarily orthonormal for a given $L$, we use the standard 
Gram-Schmidt procedure to obtain an orthonormal basis. Some technical
details ought to be mentioned here. The Metropolis algorithm
employed in our MC calculations has minimum statistical error when the weight function
behaves approximately as the wave function.  We find that it is crucial to
use several weight functions with different angular momenta $L$
in order to reduce the error to desired level.  We divide the MC calculations in 10
configurations, with the number of iterations on the order of $10^7$ for
each configuration.  Such large number of steps are required to
determine accurately the extremely small energy differences.  To reduce the
computation time for large systems, for example, $N=14$, 18 at $\nu=4/11$, we place each of
the MC configuration on a single node (dual 1GHz Intel
Pentium III processor) of a PC cluster.

\begin{figure}
\centerline{\psfig{figure=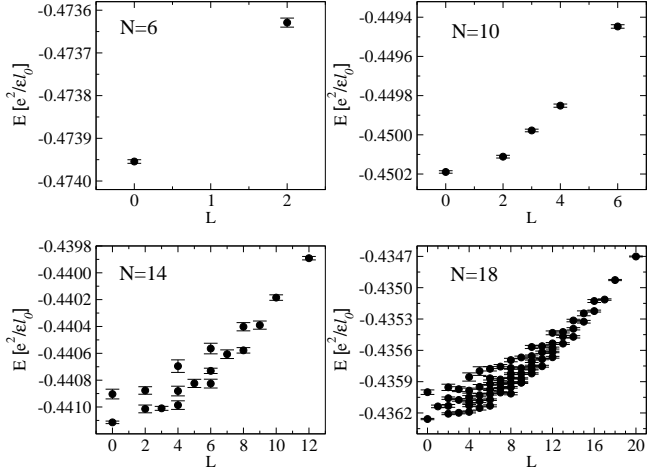,width=3.5in,angle=-90}}
\caption{The energy spectrum at $\nu=4/11$ for $N=6$, 10, 14, and 18
particles.  It is assumed that the state has partial spin polarization.
The energy per particle $E$ includes the interaction
with the uniform, positively charged background.
The quantity $l_0=\sqrt{\hbar c/eB}$ is the magnetic length at $\nu$, and
$\epsilon$ is the dielectric constant of the host semiconductor.
The error bars show the statistical uncertainty in the Monte Carlo simulation.
\label{fig1}}
\end{figure}

\begin{figure}
\centerline{\psfig{figure=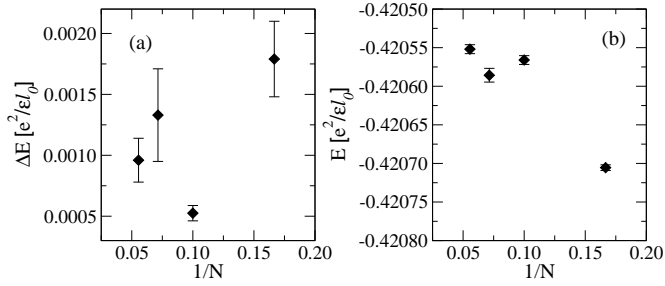,width=3.5in,angle=-90}}
\caption{Neutral excitation gap ($\Delta E$) and the ground state energy per particle
($E$) at $\nu=4/11$ as a function of $N^{-1}$, $N$ being the number of composite fermions.
\label{fig2}}
\end{figure}

Fig.~(\ref{fig1}) shows the low energy spectrum at $\nu=4/11$ for $N=6$, 10, 14, and
18, for which there are $N_{\downarrow}=$ 2, 3, 4, and 5 composite fermions
in the spin reversed CF level.
The dimension of the basis is the same as that of the lowest Landau level Hilbert
space of $N_{\downarrow}$ particles at $q^*$.
In all cases, the ground state is a uniform state with $L=0$.
From the analogy to $\nu=1/3$, it 
might be expected that the excitation spectrum contains a well defined 
branch of composite-fermion exciton \cite{Dev}, containing a single multiplet at 
$L=2,\; ...\;, N_{\downarrow}$. 
This CF exciton branch is identifiable for $N=14$ but not at $N=18$.
Nonetheless, there is a well defined gap in all cases. Fig.~(\ref{fig2})
shows the $N$ dependence of minimum energy needed 
for the creation of a CF exciton.  There are
substantial finite-size fluctuations in the value of the gap, because 
the number of spin reversed composite fermions is quite small, but 
we believe that our results indicate that the gap remains finite in
the thermodynamic limit, producing incompressibility at 4/11.
We will use the gap of the largest system studied as a rough estimate of the 
thermodynamic gap.  The next fraction
we consider is $\nu^*=1 (\uparrow) + 2/5 (\downarrow)$, corresponding to a
partially polarized state at $\nu=7/19$.  As Fig.~(3) shows, the state here is
also incompressible.  A thermodynamic extrapolation for the gap is not possible 
for $\nu=7/19$, for we have only two results, but 
the gap for $N=18$ is taken as an estimate of the thermodynamic limit.

The gaps for $\nu=4/11$ and $\nu=7/19$ are estimated to be  
$\sim 0.001 e^2/\epsilon l_0$ and $\sim 0.0008 e^2/\epsilon l_0$, which are roughly 
two orders of magnitude smaller than the gaps at $\nu=1/3$ and $\nu=2/5$, 
$0.1 e^2/\epsilon l_0$ and $0.055 e^2/\epsilon l_0$, respectively (for the 
model neglecting transverse thickness).  
Here, $l_0=\sqrt{\hbar c/eB}$ is the magnetic length at $\nu$, $B$ is 
the external magnetic field, and $\epsilon$ is the dielectric constant 
of the host material.  The gap value at $\nu=4/11$ is consistent with that quoted in
Ref.~\onlinecite{Park2}. 
It is noted that the gaps are not affected by the Zeeman energy, so long as the
partially polarized state is the ground state, because the low energy excitations of these 
states do not involve any spin reversal.  (The Zeeman energy is much higher, for typical 
experimental parameters, than the energy scales considered in this work, making 
spin flip excitations irrelevant to the low-energy physics.)
Fig.~(\ref{fig2}) estimates the thermodynamic limit of the ground
state energy at $\nu=4/11$ 
to be $\sim -0.42054 e^2/\epsilon l_0$, which also is in good accord with the
value calculated earlier\cite{Park2}.

The smallness of the gaps for the higher order FQHE states
confirms that the interaction between the composite fermions in higher CF levels
is exceedingly weak compared to the
Coulomb interaction between {\em electrons} that governs the gaps at $\nu=1/3$ and 2/5.
It is remarkable that the composite fermion theory is capable of capturing such subtle 
quantitative physics, and that 
experiments have come to a stage where higher order FQHE states
are now beginning to reveal themselves.

Table ~I gives the overlaps of the ground state with the 
wave function of Eq.~(\ref{incom}). 
The overlaps are fairly large, confirming that the trial wave 
function of Eq.~(\ref{incom}) describes the ground state effectively; in other words,
the physics of the FQHE at $\nu=4/11$ and $\nu=7/19$ is related to the 
$\nu_{\downarrow}=1/3$ and $\nu_{\downarrow}=2/5$ FQHE in the spin reversed sector.

\begin{figure}
\centerline{\psfig{figure=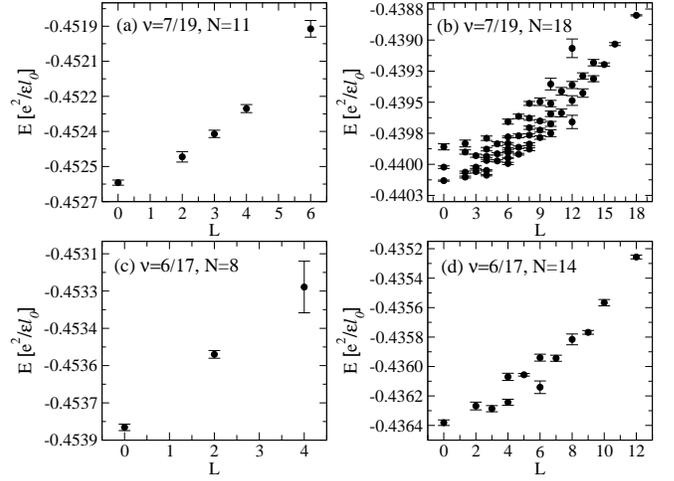,width=3.5in,angle=-90}}
\caption{The energy spectrum at $\nu=7/19$ for $N=11$ and $N=18$
and at $\nu=6/17$ for $N=8$ and $N=14$ 
for a partially spin polarized system.
\label{fig3}}
\end{figure}

We have also investigated the possibility of partially polarized QHE at 
$\nu=6/17$, which maps into $\nu^*=1(\uparrow)+1/5(\downarrow)$ of composite fermions.
The theoretical spectra for 8 and 14 particles at 
$Q_{6/17}=(17 N-22)/12$, shown in Fig.~(\ref{fig3}), provide an indication of  
an incompressible state here as well.  Surprisingly, the gap for $N=14$ is approximately 
$\sim 0.0014 e^2/\epsilon l_0$, which is of the same order as the $\nu=4/11$ gap.
However, the system sizes are effectively very small,  as can be seen by the fact 
that there are only one or two basis functions at each angular momentum, 
which prevents us from making a more reliable assertion regarding the presence of 
incompressibility at $\nu=6/17$.  At the moment, we are unable to get sufficient 
accuracy at the next particle number ($N=20$).

Our study thus predicts that partially polarized higher order FQHE at fractions of the 
form $\nu=\frac{3m+1}{8m+3}$ 
should be possible in an appropriate range of Zeeman energy and temperature.   
The temperature scale set by the gap is on the order of 150 mK (at $B=10$T) 
for GaAs, which is an upper limit because corrections due to finite thickness and 
disorder are expected to suppress the gap substantially.  
It is at present not possible to 
ascertain theoretically the relevant Zeeman energy range, for lack 
of a quantitative understanding of the fully polarized 
states at these filling factors.  The modifications due to finite transverse
thickness, not included above, are also of relevance.

Similar considerations may also be useful for the QHE-like features seen previously 
at $\nu=7/11$ in very low density samples \cite{Goldman}.
The relevance of our results to the Raman experiment \cite{Pinczuk} in 
the filling factor range $2/5 \geq \nu \geq 1/3$, where the level structure of 
composite fermions is observed, also deserves further investigation.

We thank Kwon Park and Vito Scarola for discussions and helpful comments on 
the manuscript.  This work was supported in part by the National Science
Foundation under Grant No. DMR-0240458.  We are grateful to the
High Performance Computing (HPC) group led by V. Agarwala, J. Holmes, and J. Nucciarone,
at the Penn State University ASET (Academic Services and Emerging Technologies) for
assistance
and computing time with the LION-XE cluster, and
acknowledge NSF DGE-9987589 for computer support.

\pagebreak

\begin{table}
\begin{tabular}{ccccccc}
      $\nu$    & $ \nu^*$ &  $N$   &  $Q$    & $N_{\downarrow}$ &   $q^*$ & overlap (\%) \\
\hline
    $\frac{4}{11}$     &    $1+\frac{1}{3}$ &   6    &   6.5   &     2     &    1.5  & 100 \\
               &          &   10   &   12.0  &     3     &    3.0  &  100 \\
               &          &   14   &   17.5  &     4     &    4.5  &   90.1  \\
               &          &   18   &   23.0  &     5     &    6.0  &   98.8  \\
\hline
    $\frac{7}{19}$     &    $1+\frac{2}{5}$ &   11   &   13.0  &     4     &    3.0  &  100 \\
               &          &   18   &   22.5  &     6     &    5.5  &   99.0 \\
\end{tabular}
\caption{The parameters for the 
systems studied in this work. $Q$ and $q^*$ are the monopole strengths for electrons
and composite fermions; $\nu$ and $\nu^*$ are filling factors for electrons
and composite fermions; $N$ is the total number of composite fermions and 
$N_{\downarrow}$ is the number of composite fermions  
in the reversed spin sector. The last column shows the overlap of the ``ground state"
at $\nu=4/11$ and $\nu=7/19$ determined by diagonalization  
in the truncated basis (see text for details)
with the wave function derived from the $\nu^*_{\downarrow}=1/3$ 
or $\nu^*_{\downarrow}=2/5$ incompressible state according to Eq.~(\ref{incom}). 
The overlap is 100\% when there is only a single uniform basis state.}
\end{table}

\begin{references}


\bibitem{Pan} W. Pan {\em et al.},
Phys. Rev. Lett., {\bf 90}, 016801 (2003);  Int. J. Mod. Phys. 
{\bf 16}, 2940 (2002).

\bibitem{Pinczuk}  Irene Dujovne {\em et al.},
Phys. Rev. Lett. {\bf 90}, 036803 (2003). 

\bibitem{Mandal} S.S. Mandal and J.K. Jain, Phys. Rev. B {\bf 66}, 155302 (2002).

\bibitem{Park2} K. Park and J.K. Jain, Phys. Rev. B {\bf 62}, R13274 (2000).

\bibitem{Quinn} A. W\'{o}js and J.J. Quinn, Phys. Rev. B {\bf 61},
2846 (2000).

\bibitem{Lee} S.Y. Lee, V.W. Scarola, and J.K. Jain, Phys. Rev. B {\bf 66}, 085336 (2002).

\bibitem{Tsui} D.C. Tsui, H.L. Stormer, and A.C. Gossard, Phys
Rev. Lett. {\bf 48}, 1559 (1982).

\bibitem{Klitzing}  K. v. Klitzing, G. Dorda, and M. Pepper, Phys. Rev. 
Lett. {\bf 45}, 494 (1980).

\bibitem{Jain}  J.K. Jain, Phys. Rev. Lett. {\bf 63}, 199 (1989);
Physics Today {\bf 53}(4), 39 (2000).

\bibitem{Spin_ex}  R.R. Du {\em et al.},
Phys. Rev. Lett. {\bf 75}, 3926 (1995);
Phys. Rev. B {\bf 55}, R7351 (1997); R.J. Nicholas {\em et al.}
Semicond. Sci. Technol. {\bf 11}, 1477 (1996); I.V. Kukushkin,
K. v. Klitzing, and K. Eberl, Phys. Rev.  Lett. {\bf 82}, 3665 (1999).

\bibitem{Wu} X.G. Wu, G. Dev, and J.K. Jain, Phys. Rev. Lett. {\bf 71}, 153
(1993).

\bibitem{Park} K. Park and J.K. Jain, Phys. Rev. Lett.  {\bf 80}, 4237 (1998).

\bibitem{Laughlin} R.B. Laughlin, Phys. Rev. Lett.  {\bf 50}, 1395 (1983). 

\bibitem{Halperin} B.I. Halperin, Helv. Phys. Acta {\bf 56}, 75 (1983).

\bibitem{Monopole} For further details on the spherical geometry
and monopole harmonics, see
T.T. Wu and C.N. Yang, Nucl. Phys. B {\bf 107}, 365 (1976);
F.D.M. Haldane, Phys. Rev. Lett. {\bf 51}, 605 (1983).

\bibitem{JK} J.K. Jain and R.K. Kamilla, Int. J. Mod. Phys. B {\bf 11}, 2621
(1997); Phys. Rev. B {\bf 55}, R4895 (1997).

\bibitem{Dev} G. Dev and J.K. Jain, Phys. Rev. Lett. {\bf 69}, 2843 (1992).

\bibitem{Goldman} V.J. Goldman and M. Shayegan, Surface Science {\bf 229}, 10 (1990).

\end{references}
\end{document}